\documentclass[prl,twocolumn,showpacs,epsf]{revtex4}
\usepackage{graphicx}
\usepackage{color}
\begin{document}
\draft
\title
{Precision Measurement of the Decay Rate of $^7$Be in Host
Materials}

\author{Y.~Nir-El$^1$, G.~Haquin$^1$, Z.~Yungreiss$^1$, M.~Hass$^2$, G.~Goldring$^2$,
S.K.~Chamoli$^2$, B.S.~Nara Singh$^2$, S.~Lakshmi$^2$,
U.~K\"{o}ster$^{3,4}$, N.~Champault$^3$, A.~Dorsival$^3$,
G.~Georgiev$^{3,5}$, V.N~Fedoseyev$^3$, B.A.~Marsh$^6$, D.
Schumann$^7$, G. Heidenreich$^8$, S. Teichmann$^8$ \\
}

\affiliation{ {{$^1$ Radiation Safety Division, Soreq Nuclear Research Centre, Yavne, Israel}\\
{$^2$ Department of Particle Physics, Weizmann Institute of Science,
Rehovot, Israel}\\ {$^3$ ISOLDE, CERN,
Geneva, Switzerland}\\ {$^4$ Institut Laue Langevin, Grenoble, France}\\
{$^5$ CSNSM, CNRS/IN2P3; Univ Paris-Sud, ORSAY-Campus, France}\\
{$^6$ Physics Department, University of Manchester, Manchester, UK}\\
{$^7$ Laboratory of Radiochemistry, Paul Scherrer Institute,
Villigen, Switzerland}\\
{$^8$ Target Facilities and Active Technique Section, Accelerator
Division, Paul Scherrer Institute, Villigen, Switzerland} }}
\date{\today}
\begin{abstract}
A controlled and precise determination of the cross-sections of the
fusion reactions $^{7}$Be(p,$\gamma$)$^{8}$B and
$^{3}$He($^{4}$He,$\gamma$)$^{7}$Be, which play an important role in
determining the solar neutrino flux, necessitates the knowledge of a
precise value of the electron-capture half-life of $^{7}$Be. This
half-life may depend on the material hosting the $^7$Be atoms via
small modifications of the electron density around the $^{7}$Be
nucleus. In this brief communication we report on the measurement of
$^{7}$Be implanted in four materials: copper, aluminum, sapphire and
PVC. The four results are consistent with a null host dependence
within two standard deviations and their weighted average of
53.236(39)d agrees very well with the adopted value in the
literature, 53.22(6)d. The present results may exhibit a slight
(0.22\%) increase of the half-life at room temperature for metals
compared to insulators that requires further studies.
\end{abstract}


\maketitle The decay rate of radioactive nuclei that undergo orbital
Electron Capture (EC) depends on the properties of the atomic
electron cloud around the nucleus. Hence, EC may exhibit varying
decay rates if the nucleus is implanted into  host materials with
different properties of their corresponding electron clouds. The
first suggestion of this effect in $^{7}$Be, which is the lightest
nucleus that decays by EC, and reports of experiments trying to
investigate this phenomenon,  have been presented by Segr\`{e} et
al.~\cite{Segre,SegreWie,Lein}. This effect has been qualitatively
attributed in the past to the influence of the electron affinities
of neighboring host atoms ~\cite{Das}. The electron density of the
$^{7}$Be atom in a high-electron affinity material such as gold is
decreased via the interaction of its 2s electrons with the host
atoms, resulting in a lower decay rate (longer half-life). Recently,
the life-time modification has been suggested to stem from
differences of the Coulomb screening potential ~\cite{Kettner}
between conductors and insulators (see below).

Several experimental and theoretical investigations were conducted
during recent years to study the host material effect on the decay
rate of $^{7}$Be ~\cite{Das,Ray,Norman,Ray1,Ohtsuki,Ray2}, with a
somewhat confusing scattering of experimental results. It was found
that the half-life of $^{7}$Be encapsulated in a fullerene C$_{60}$
cage and $^{7}$Be in Be metal is 52.68(5) and 53.12(5) days,
respectively, amounting to a difference of 0.83(13)\%
\cite{Ohtsuki}. A smaller effect of $\approx$0.2\% was measured for
the half-life 53.64(22) d of $^{7}$Be in C$_{60}$ and 53.60(19) d of
$^{7}$Be in Au ~\cite{Ray2}. A recent theoretical evaluation shows
that short and long half-lives 52.927(56)d and 53.520(50)d were
measured for $^{7}$Be in Al$_{2}$O$_{3}$ and $^{7}$Be in (average of
BeO, BeF$_{2}$ and Be(C$_{5}$H$_{5}$)$_2$), respectively
~\cite{Das}. These results yield the magnitude of the effect to be
as large as 1.1$\%$. Another experimental investigation has shown
that the half-life increases by 0.38$\%$ from $^{7}$Be in graphite,
53.107(22) d, to $^{7}$Be in Au, 53.311(42) ~\cite{Norman}, while a
very recent investigation ~\cite{Limata} has seen no effect to
within 0.4\%. The great interest in this phenomenon for $^7$Be
arises also from the need to explore the possible contribution of
the half-life of $^{7}$Be to the measurement of the cross section of
the two fusion reactions, $^{7}$Be(p,$\gamma$)$^{8}$B and
$^{3}$He($^{4}$He,$\gamma$)$^{7}$Be, that play an important role in
determining the solar neutrino flux ~\cite{Baby,Nara}.

The present work has been undertaken in order to probe this
phenomenon yet again in an experimental approach that takes full
advantage of the experience gained in measuring implanted $^{7}$Be
activity in a controlled and precise manner for cross section
determinations of the solar fusion reactions mentioned above
~\cite{Baby,Nara}. As a demonstration of the quality of the
$\gamma$-activity measurement, we cite the results of ~\cite{Baby,
Baby1} where two independent determinations of the $\it{absolute}$
activity of $^7$Be, at the Soreq laboratory and at Texas A$\&M$
University, were in excellent agreement to within 0.7$\%$.  The same
setup has also been used for determining the $^7$Be activity ensuing
from the $^{3}$He($^{4}$He,$\gamma$)$^{7}$Be reaction ~\cite{Nara}.
We report the measurement of the half-life of $^{7}$Be implanted in
four host materials: copper, aluminum, aluminum oxide (sapphire -
Al$_{2}$O$_{3}$) and PVC (polyvinyl chloride - [C$_2$H$_3$Cl]$_n$]).

The primary source of $^{7}$Be for implantation was a graphite
target, from the Paul Scherrer Institute (PSI), used routinely for
the production of $\pi$ mesons ~\cite{Heid}. Many spallation
products are accumulated in the target, including $^7$Be. Graphite
material from the PSI meson production target was placed in an
ion-source canister and was brought to ISOLDE (CERN); $^7$Be was
extracted at ISOLDE by selective ionization using a resonance laser
ion source. Direct implantation of $^{7}$Be at 60 keV in the host
material was subsequently followed. A detailed description of the
extraction and implantation of $^7$Be at ISOLDE is provided in
detail in Refs. ~\cite{Baby1,Kost}. This procedure facilitated a
precision measurement of the cross section of the reaction
$^{7}$Be(p,$\gamma$)$^{8}$B. The implantation spot was defined by a
2 mm collimator positioned at close proximity to the target for the
Cu sample and a 5 mm collimator for the other samples. This small
change in the ensuing counting geometry has been well investigated
for the measurement of $^7$Be activity ~\cite{Nara} and does not
affect the results in any significant manner. The implantation
process provided full control of the spot composition ($^{7}$Be;
$^{7}$Li) as well as a radial and depth profiles. For earlier
implantations, at a density of $^7$Be in Cu far exceeding that of
the present experiment, the spot was found to be robust and the
$^{7}$Be inventory in the spot was stable ~\cite{Baby1,Hass},
excluding naturally radioactive decay. The copper, aluminum and PVC
host material targets consisted of disks of 12 mm diameter and 1.5
mm thickness, while the sapphire target was a square of 10.2 mm x
10.2 mm. The median implantation depth of $^{7}$Be into these
materials has been estimated using the SRIM code ~\cite{Srim} and
found to be 12, 24, 470 and 37 $\mu$m, respectively, i.e. all
implantation depths were well below the surface.

Reproducible counting geometry of the $^{7}$Be samples was achieved
by attaching them to plastic holders which were mounted precisely on
the detector endcap. Also attached to the holders was a $^{133}$Ba
source ($\it T$$_{1/2}$ = 3841(7) d ~\cite{Tuli}) to correct for
variations in the performance of the detection system (geometry,
detection efficiency). In a separate set of measurements, an
external $^{137}$Cs source ($\it T$$_{1/2}$ = 30.03(5)y
~\cite{Tuli}) was attached to a similar holder and thus used to
estimate the reproducibility of source position by successive
mountings of that holder.

Gamma-ray spectra were acquired by a p-type coaxial HPGe detector of
63.7$\%$ relative detection efficiency, 1.78 keV energy resolution
(FWHM) and 81.6 peak/Compton ratio, all specified at the 1332.5 keV
gamma-ray of $^{60}$Co. The detector was enveloped by a 5.1 cm
mercury cylinder and placed within a 10.2 cm thick lead shield. The
electronic train following the detector consisted of standard units,
followed by a 8192 channels multichannel analyzer.

$^{7}$Be decays by EC to the ground and first excited state of
$^{7}$Li at 477.6 keV. The branching ratio to $^{7}$Li*(477.6) is
10.44(4)\% and the adopted half-life is 53.22(6) d ~\cite{Trilley}.
This general-use value of the half-life was intended by the
evaluator, the late R. Helmer, to be valid for Be and BeO samples
and adequate for various chemical forms ~\cite{Trilley}.

Measurements of all $^{7}$Be samples were repeated approximately
every two weeks and the total decay time between the beginnings of
first and last measurements of a sample was 171.1 (copper), 146.0
(PVC), 163.0 (aluminum) and 180.6 d (sapphire). Measurements were
stopped when the statistical uncertainty of  the 477.6 keV peak
reached 0.15 to 0.10 $\%$. At the beginning of the measurements, the
activities of each sample were: 1900, 1800, 2600 and 3700 Bq,
respectively and a typical counting duration was 16 hours.

Since the extracted half-life values may depend slightly on the
analysis method used to compute peak areas, we describe in some
detail one such procedure of data analysis that was used to extract
the present results. Other analysis procedures were tried as well,
without affecting the conclusions in any significant manner.

The net number of counts in the 477.6 keV peak of $^7$Be was
calculated by summation  after the integral counts of the two scaled
broad flat windows, to the left and right of the peak, were
subtracted from the gross integral counts in the Region Of Interest
(ROI) of the peak, as can be seen in Fig. 1.

\begin{figure}[ht]
\includegraphics[width=9.0 cm]{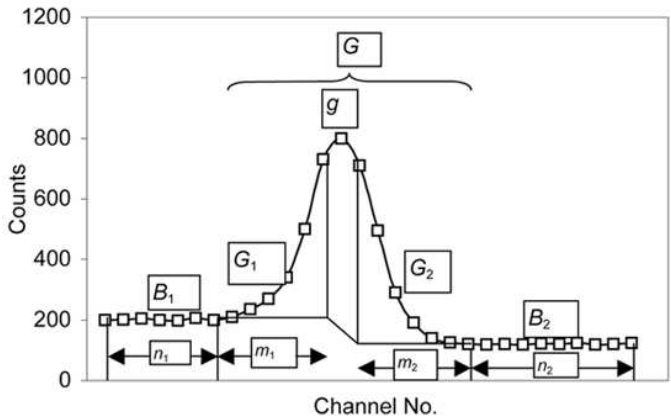}
\hfill \caption{Definition of 4 regions for the calculation of a net
peak area by the summation method. The widths are $\it n$$_1$, $\it
m$$_1$, $\it m$$_2$ and $\it n$$_2$ (in channels) and the integral
numbers of counts are $\it B$$_1$, $\it G$$_1$, $\it G$$_2$ and $\it
B$$_2$, respectively, depicting the case where at the top there is a
single channel of $\it g$ counts.} \label{Fig.1}
\end{figure}

\begin{figure}[ht]
\includegraphics[width=9.0 cm]{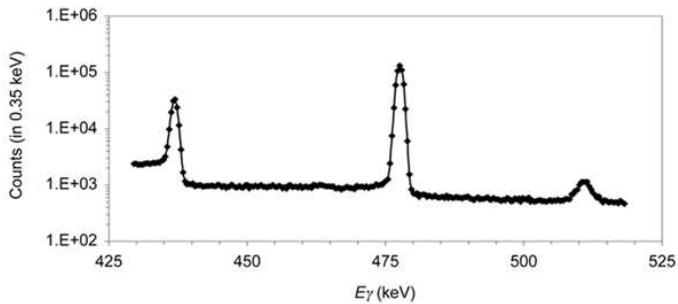}
\hfill \caption{An expanded view (in logarithmic scale) around the
$^7$Be peak at 477.6 keV. The quality of the $\gamma$ spectrum, the
two flat regions around the $^7$Be line and the absence of any
interfering $\gamma$ lines can be well discerned.} \label{Fig.2}
\end{figure}

The edges of the windows were determined to avoid overlap with the
peaks at 437.0 keV (sum of 81.0 and 356.0 keV of $^{133}$Ba) and
511.0 keV (annihilation) and the ROI was determined to include the
low- and high-energy tails of the peak (see Fig. 2).

  The gross number of counts in the ROI is given by: $G=G_1+G_2+g$
where $\it g$ is the number of counts in the channel at the top of
the peak.

The baseline to be subtracted is \begin {equation} B={m_1+0.5\over
n_1} B_1 + {m_2+0.5\over n_2}B_2\end {equation}
 The net number of counts and its standard
uncertainty are $N=G-B$ and $\sigma(N)=[G+\sigma^2(B)]^{0.5}$ where
the variance of the baseline is given by

\begin {equation} \sigma^2(B)=\left({m_1+0.5\over
n_1}\right)^2 B_1 + \left({m_2+0.5\over n_2}\right)^2B_2\end
{equation}

If the peak is symmetrical, i.e. two channels of equal height at the
top, then $\it g$ = 0 and the 0.5 terms in Eqs. (1) and (2) must be
removed.

The peak count-rate at the beginning of data acquisition
is given by
\begin {equation} R_0={\lambda N \over 1-[exp(-\lambda T)]}
\end{equation}
\noindent where $\it T$ is the acquisition Real-Time (in s) and the
decay constant (in s$^{-1}$) of $^7$Be is $\lambda={ln(2) \over
T_{1/2}}$. $\it T$$_{1/2}$  is the half-life (in s) of $^{7}$Be and
its initial value in Eq. (3) can be chosen as 53.22 d. The
contributions of the small errors on $\it T$ and $\lambda$ to the
propagated uncertainty are negligible. The standard uncertainty of
$\it R$$_{0}$ is then: $\sigma(R_o)=R_o{\sigma(N) \over N}$.
\noindent With $R^{(o)}_o$ being the count-rate at the beginning of
the first measurement ($\it t$ = 0) of a sample, the exponential
decay can be expressed by the linear relationship
$ln(R_o)=ln(R^{(o)}_o)-\lambda t$ \noindent. A weighted linear
regression of $\it ln(R$$_0$) versus $\it t$ gives the slope
$\lambda$ and the standard uncertainty $\sigma(\lambda)$. Hence,
$\it T$$_{1/2}$ and $\sigma$($\it T$$_{1/2}$) can be obtained. The
calculated value of $\lambda$ was substituted in Eq. (3), instead of
the initial value, and the linear regression was repeated.
Convergence was achieved after one iteration.  The linear fitting
procedure was examined by 3 criteria: (a) the correlation
coefficient $\it r$, (b) the reduced chi-square $\chi ^{2}$/$\nu$,
and (c) the probability $\it P$$_{\chi}$($\chi ^{2}$, $\nu$) that
any random set of $\it n$ data points would yield a value of
chi-square as large as or larger than $\chi ^2$.  The goodness of
the fit is determined by how close are the extracted criteria to the
optimal values of -1, +1 and 100\%. The number of degrees of freedom
$\nu$ is equal to $\it n$-2 for the fitting of a straight line (the
single coefficient is the slope and the constant term is the
intercept).

\begin{figure}[ht]
\includegraphics[width=9.0 cm]{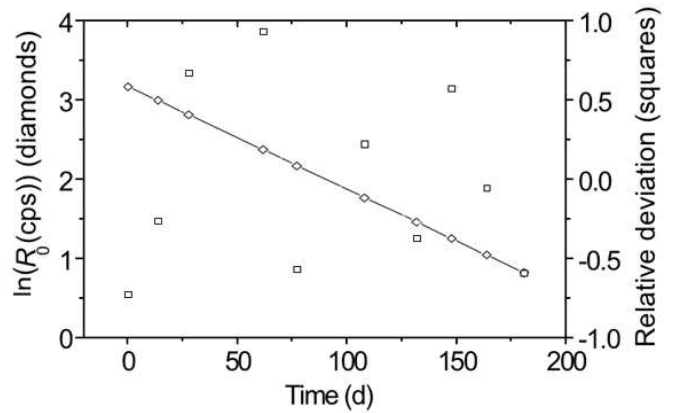}
\hfill \caption{Left axis - (diamonds): The decay curve of the 477.6
keV line of $^7$Be imbedded in the Al$_2$O$_3$ sample. The straight
line was fitted by a weighted linear regression. Right axis -
(squares): The deviation between the measured and the fitted count
rates, divided by the corresponding uncertainty, for the 10
measurements of the Al$_2$O$_3$ sample (see text).} \label{Fig.3}
\end{figure}

The decay of the peak count-rate of the Al$_{2}$O$_{3}$ sample is
shown as an example in Fig. ~\ref{Fig.3}. Uncertainties of
count-rates were of the order 0.10 to 0.16\% and express the high
precision of the measurements. Fig. ~\ref{Fig.3} shows also that the
calculated straight line fits well the measured results.
The goodness of the linear fit of the Al$_{2}$O$_{3}$ sample is
displayed in Fig. ~\ref{Fig.3}, which shows the difference between
measured and calculated (fitted) count-rates, in units of the
associated uncertainty of the measured value.

\begin{table}
\caption{The details of the half-life determinations and the linear
regression fits for the four samples. $r$ is the correlation
coefficient defined in the text.} \vspace{0.5cm}
\begin{tabular}{llllcccc}
\hline
Host&{\it n}&&&$T_{{1\over2}}(d)$&($r$+1)10$^6$&$\chi^2_\nu $&$P_\chi(\chi^2,\upsilon)$(\%)\\
Material\\
\hline
Cu&10&&&53.353(50)&1.9&0.94&48.56\\
Al&10&&&53.257(44)&0.9&0.73&66.53\\
Al$_2$O$_3$&10&&&53.180(43)&0.4&0.39&92.73\\
PVC&10&&&53.181(45)&1.7&1.26&25.88\\
\hline
\end{tabular}
\end{table}

 The geometrical uncertainty, found by the $^{137}$Cs
source to be 0.035\%, was applied to each data point in quadrature
addition to the statistical uncertainty $\sigma(R_o)$. The
analytical uncertainty 0.062\% was determined by running three
different methods to analyze a peak area.  This uncertainty was
added in quadrature to the provisional uncertainty as calculated by
the linear regression.

\begin{figure}[!h]
\includegraphics[width=9.0 cm]{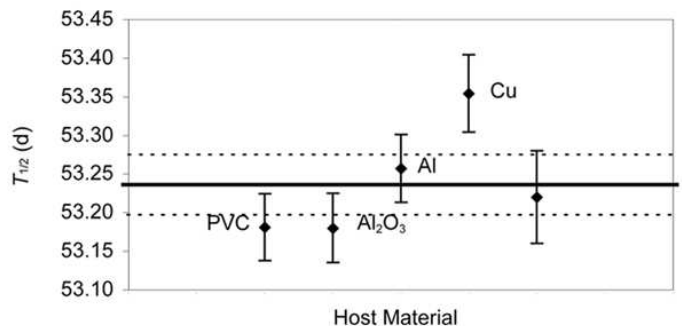}
\hfill \caption{The half life of $^7$Be in 4 host materials. The
solid line represents the weighted average and the broken lines
correspond to a $\pm$1$\sigma$ interval. Also shown is the adopted
value in the literature ~\cite{Trilley}} \label{Fig.4}
\end{figure}

The weighted average of the measured half-lives of the four host
materials, presented in Table I and in Fig. ~\ref{Fig.4}, is
53.236(39) d, which agrees well with the adopted value 53.22(6) d
~\cite{Trilley}, with no account being taken of the host material.

%\begin{figure}[h]
%\includegraphics[width=9.0 cm]{Fig.3.eps}
%\hfill \caption{} \label{Fig.3}
%\end{figure}

Even though the statistical test of the present data supports a null
effect within $\pm$ 1$\sigma$, the results of Fig. ~\ref{Fig.4} may
indicate a slight positive trend of the half-life versus the
electron affinity, where a host material with high-electron affinity
such as copper (conductor), exhibits a longer half-life, compared to
a lower electron affinity material such as aluminum oxide
(insulator).

A different possible interpretation of the life-time results follows
a recent observation by Wang et al. ~\cite{Wang} of an approximately
1$\%$ increase in the lifetime of $^7$Be in metallic vs. insulator
environments at low temperature. This change is consistent with the
Debye screening model ~\cite{Kettner} that has been successfully
used to explain the screening potential for nuclear reactions at
very low beam energies. The average life times for the two
insulators (PVC and Al$_2$O$_3$) and the two metals (Cu and Al) are
53.180(31) d and 53.299(33) d, respectively, a difference of
0.22$\%$. Indeed, the trend of the present data is in basic
agreement with the temperature dependence of the screening model, as
well as with the results of ~\cite{Limata}. A further investigation
of such a small trend and its detailed temperature dependence is
clearly called for.

We thank the PSI technicians Pedro Baumann and Alfons Hagel for the
target handling and the ISOLDE staff for their help. The work has
been supported in part by the Israel Science Foundation and the
EU-RTD project TARGISOL (contract HPRI-CT-2001-50033). We
acknowledge the support of the ISOLDE Collaboration.

\end{document}